\documentclass[11pt]{article}

\usepackage{graphicx}
\usepackage{epstopdf}
\usepackage{natbib}
\usepackage{xspace}
\usepackage{xcolor}
\usepackage{amsmath, amssymb}

\usepackage{geometry}
\newcommand\etc{\textit{etc.}\xspace}
\newcommand\eg{\textit{e.g.}\xspace}
\newcommand\ie{\textit{i.e.}\xspace}

\renewcommand\Re{\mbox{$\mathrm{Re}$}}
\renewcommand\Im{\mbox{$\mathrm{Im}$}}

\def \B {\mbox{${\boldsymbol B}$}}

\def \J {\mbox{${\boldsymbol j}$}}

\def \Rh {\hat{\mathbf R}}
\def \phih {\hat{\boldsymbol{\phi}}}
\def \zh {\hat{\boldsymbol z}}
\providecommand\bnabla{\boldsymbol{\nabla}}

\def \eb {\mathbf{e}}

\def \Jac {J}

\usepackage{unicode}
\newcommand{\csso}{\fontencoding{LECO}\selectfont\char215}
\newcommand{\iotaslash}{\hspace*{0.1em}\iota\hspace*{-0.45em}\text{\csso}}

\newcommand{\poiss}[2]{\left\{ #1, #2 \right\}}
\newcommand{\poissRZ}[2]{\left\{ #1, #2 \right\}_{RZ}}
\newcommand{\poissRZo}[2]{\left\{ #1, #2 \right\}_{R_0Z_0}}
\providecommand\bnabla{\boldsymbol{\nabla}}

\providecommand\bphi{\boldsymbol{\phi}}

\usepackage{authblk}

\title{Quasi-axisymmetric magnetic fields: weakly non-axisymmetric case in a vacuum}
\author[1]{G.G. Plunk\thanks{gplunk@ipp.mpg.de}}
\author[1]{P. Helander}
\affil[1]{Max-Planck-Institut f\"ur Plasmaphysik, EURATOM Association, 17491 Greifswald, Germany}

\begin{document}
\maketitle

\begin{abstract}
An asymptotic expansion is performed to obtain quasi-axisymmetric magnetic configurations that are weakly non-axisymmetric.  A large space of solutions is identified, which satisfy the condition of quasi-axisymmetry on a single magnetic flux surface, while (non-axisymmetric) globally quasi-axisymmetric solutions are shown to not exist, agreeing with the conclusions of previous theoretical work.  The solutions found are shown to be geometrically constrained at low aspect ratio or high toroidal period number.  Solutions satisfying the more general condition of omnigeneity (generalized quasi-axisymmetry) are also shown to exist, and it is found that quasi-axisymmetric deformations can be superposed with an omnigenous solution, while preserving the property of omnigeneity, effectively extending the space of ``good'' configurations.  A numerical solution of the first order quasi-axisymmetry problem is demonstrated and compared with solutions found with a widely used MHD equilibrium solver, independently verifying that quasi-axisymmetry is satisfied at the appropriate order.  It is thereby demonstrated that approximately quasi-axisymmetric solutions can be directly constructed, \ie without using numerical search algorithms.
\end{abstract}

\section{Introduction}

3D MHD equilibrium solutions with topologically toroidal magnetic flux are conventionally specified by the plasma current, pressure profile and the shape of a flux surface \citep{bauer-betancourt-garabedian}.  In a vacuum, merely imposing the surface shape is sufficient; this corresponds to a Neumman boundary condition, since the magnetic field, which is then equal to the gradient of a scalar field, has no component perpendicular to the surface, and so the normal derivative of the scalar field is zero.  It seems plausible that another kind of differential constraint might be substituted at the boundary, instead of surface shape, to constitute an alternative boundary condition (or an alternative ``problem specification'').  The condition of quasi-symmetry, a symmetry of the magnetic field strength expressed in Boozer angles \citep{nuehrenberg-zille}, is one such possibility.  This property ensures that the magnetic field confines collisionless particle orbits.  In fact, it was argued by \citet{garren-boozer-2} that, although it is generally not possible to satisfy quasi-symmetry across the entire volume, it should be possible to do so exactly on one surface, since there is a sufficient amount of freedom in the solution.  However, a proof of the existence of such solutions has not been given, nor is it known how many such solutions exist.  Our understanding of such issues is even more incomplete for the general class of ``omnigenous'' equilibria, which have good particle confinement without (necessarily) having quasi-symmetry.

A subclass of quasi-symmetry, known as quasi-axisymmetry (QAS) \citep{nuehrenberg-QAS}, is particularly interesting, as it can be thought of as a generalization of axisymmetry.  It has been suggested that QAS configurations might be accessed by non-axisymmetric shaping of tokamak equilibria \citep{boozer-ITER}.  Practically speaking, non-axisymmetric shaping is a possible avenue for mitigating major problems faced with tokamaks, related to stability and disruptions \citep{ku-boozer}.  The stabilizing influence of non-axisymmetric shaping has been verified experimentally in recent years \citep{CTR}, but the shaping was not designed to preserve QAS.  In light of the success of optimized stellarator design, \ie with the Helically Symmetric Experiment and Wendelstein 7-X, it seems realistic to expect that stabilization by non-axisymmetric shaping of tokamaks could be achieved while also preserving the good confinement of particle orbits.  This further motivates us to reconsider the theory of QAS magnetic fields.

In the present work, we perform an expansion about axisymmetry, establishing the existence of QAS magnetic fields (\ie satisfying QAS on a single surface), characterizing the space of these solutions, and relating them to the more general class of ``omnigenous'' solutions.  A numerical method to explicitly find QAS solutions is also provided.  QAS is different from other forms of quasi-symmetry, since there is a (known) case (namely true axi-symmetry) where the QAS condition is exactly satisfied globally, and for this reason it has been suggested that QAS equilibria may be obtained by continuous deformation of tokamak equilibria .  (Equilibria with helical or cylindrical symmetry are globally quasi-symmetric, but such equilibria do not close toroidally.)  We find that QAS can indeed be satisfied at a single surface, to all orders in the expansion, lending support to the conclusion of \citet{garren-boozer-2}, though not establishing it exactly (\ie non-perturbatively).  Note here that QAS is satisfied on the outer surface but the magnetic field solution is global, satisfying the magnetostatic equations all the way to the magnetic axis.  This approach can be contrasted with local equilibrium theory, which solves the magnetostatic equilibrium equations only in the region close to a given flux surface \citep{hegna-2000, boozer-2002}.

In our expansion, the deviation from axisymmetry is controlled by the (arbitrary) small parameter $\epsilon$.  We consider the vacuum case, because it is significantly simpler than the non-vacuum case; it also has the convenient feature that any obtained rotational transform can be attributed entirely to 3D shaping, since none can arise from the zeroth-order axisymmetric solution.  

Our findings can be summarized as follows.  Due to axi-symmetry at zeroth order, the linear first order equations have an ignorable coordinate, and thus a free parameter, the toroidal mode number $N$.  We treat this as a parameter of first-order deformation, and note that it can then be interpreted as the number of field periods of the stellarator, since higher order contributions only involve harmonics of this fundamental mode number.  From the first order equations, a single linear 2nd order PDE for a single scalar field is obtained, whose solution completely determines the magnetic field.  QAS is expressed as a simple differential constraint, and it is demonstrated that it cannot be satisfied globally, except for the trivial case where axisymmetry is preserved by the perturbation.  The constraint of QAS can, however, be applied on a single surface as a non-standard boundary condition for the PDE, in which case the problem takes the form of an ``oblique derivative'' problem.  The solution of this mathematical problem is generally non-unique, but for our case (a homogeneous elliptic PDE), exactly two linearly independent solutions are guaranteed to exist.  We conclude that there is a certain freedom in these single-surface QAS solutions:  (1) The freedom of an arbitrary surface shape for the zeroth order axisymmetric flux surfaces; (2) the toroidal mode number $N$ of the deformation (toroidal field period); and (3) two complex amplitudes for the independent solutions of the oblique derivative problem.

At second order, the magnetostatic equation, with the QAS condition, is again reformulated as an oblique derivative problem, but in this case the system is non-homogeneous.  The problem is known to have solutions, but the number of solutions cannot generally be predicted.  The externally induced rotational transform (\ie due to the non-axisymmetric deformation) is obtained at second order (but in terms of first order quantities), \ie it scales as $\epsilon^2$ -- this is the amount that can be attributed to the non-axisymmetric deformation.  At third and higher order, the problems takes a form similar to the second order problem, and therefore a solution is guaranteed to exist at each order.  We conclude that QAS can be satisfied at all orders in the expansion.

At the surface where QAS is satisfied, a local magnetostatic condition is derived, from which a general conclusion can be drawn about QAS magnetic fields, namely that the non-axisymmetric perturbation in the field is confined to the inboard side at low aspect ratio, and the effect is amplified at high toroidal mode number.  This may be a useful consideration in the design of QAS devices.  Finally, we observe that, as an alternative to QAS, the condition of omnigeneity (also referred to as ``generalized quasi-symmetry'' \cite{landreman}) can be imposed on the field strength $B_0 + \epsilon B_1$ as a function of Boozer angles on a single magnetic surface.  It is noted that the QAS solution represents a homogeneous solution to this general problem, and therefore represent an arbitrary freedom in the solution.

The paper is organized as follows.  In Sec.~\ref{prelim-sec}, the basic equations, assumptions, and notation are introduced.  In Sec.~\ref{inverse-formulation-sec}, the problem formulation is stated, and the expansion about axi-symmetry is performed.  The main theoretical results are presented here, although some details are contained in appendices.  In Sec.~\ref{numeric-sec}, a numerical solution of the first order QAS problem is demonstrated, and a comparison with the widely-used VMEC equilibrium solver \citep{Hirshman} is made to confirm the validity of the solutions.  In Sec.~\ref{gen-qas-sec}, the extension of the method to treat generalized QAS (omnigeneity) is discussed.  Sec.~\ref{conclusion-sec} contains further discussion and conclusions.

\section{Preliminaries}\label{prelim-sec}

The MHD equilibrium equations are

\begin{eqnarray}
&\bnabla\times\B = \mu_0\J,\label{ampere-eqn}\\
&\bnabla\cdot\B = 0,\label{div-B-eqn}\\
&\J\times\B = \bnabla \psi \frac{d p}{d\psi},\label{force-balance-eqn}
\end{eqnarray}

\noindent We assume topologically toroidal flux surfaces, labeled by $\psi$, such that $p = p(\psi)$.  To solve these equations, we use a similar approach as previous works \citep{garren-boozer-1, garren-boozer-2, hegna-2000, boozer-2002, Weitzner}.  Boozer angles \citep{booz-coords-1981} are denoted $\theta$ and $\varphi$.  The contravariant form of $\B$ is written

\begin{equation}
\B_{\mathrm{con}} = \bnabla\psi \times \bnabla\theta - \iotaslash \bnabla \psi \times \bnabla \varphi,\label{B-contra-eqn}
\end{equation}

\noindent where $\iotaslash$ is the rotational transform, and $2\pi \psi$ is the toroidal magnetic flux.  This form of $\B$ satisfies zero divergence and assumes flux-surface geometry.  The covariant form is written

\begin{equation}
\B_{\mathrm{cov}} = G(\psi) \bnabla\varphi + I(\psi) \bnabla\theta + K(\psi, \theta, \varphi) \bnabla \psi,\label{B-covar-eqn}
\end{equation}
where $G$ and $I$ are poloidal and toroidal currents, respectively.  This form is a consequence of $\J \cdot \bnabla \psi = 0$ (\ref{force-balance-eqn}), and Ampere's law (\ref{ampere-eqn}), which itself implies $\bnabla\cdot\J = 0$.  See \eg \citet{helander-review}.  

The basic strategy to find an equilibrium is to assert $\B_{\mathrm{con}} = \B_{\mathrm{cov}}$ together with force balance (\ref{force-balance-eqn}), relying on the fact that these forms of the magnetic field incorporate Eqns.~\ref{ampere-eqn} and \ref{div-B-eqn} as well as the assumption of magnetic flux surfaces.  Either the magnetic coordinates $\psi$, $\theta$, and $\varphi$ can be considered as unknown functions of spatial coordinates (``direct formulation"), or the coordinate mapping ${\boldsymbol x}(\psi, \theta, \varphi)$ can be considered as an unknown function of magnetic coordinates (``inverse formulation'').  The direct formulation is appropriate for the axisymmetric problem, where symmetry is defined relative to the geometric toroidal angle, while the inverse formulation is appropriate for higher orders in the expansion, where axisymmetry is broken and the condition of QAS can be enforced as a symmetry in the Boozer toroidal angle $\varphi$.

In what follows we assume vacuum conditions, $dG/d\psi = 0$, and $I = K = 0$.  The vacuum case has the curious feature that flux surfaces are not uniquely defined for the zeroth-order axisymmetric solution, since the field has no rotational transform (actually, we will find the rotational transform in our expansion to also be zero at first order).  However, the non-axisymmetric perturbation introduces a non-zero rotational transform, which then makes the surfaces unique.

\section{Quasi-axisymmetry near axisymmetry}\label{inverse-formulation-sec}

In the following, to more easily apply the condition of QAS, we employ the ``inverse'' formulation, where unknown of the theory is the coordinate mapping ${\boldsymbol x}(\psi, \theta, \varphi)$, and the magnetostatic equations are written in terms of various derivatives, denoted as $\eb_1 = \partial {\boldsymbol x}/\partial \psi$, $\eb_2 = \partial {\boldsymbol x}/\partial \theta$, and $\eb_3 = \partial {\boldsymbol x}/\partial \varphi$.  These equations can be translated into equations involving the metrics via the usual identities (reviewed in Appx.~\ref{geometry-appx}).  The equation $\B_{\mathrm{con}} = \B_{\mathrm{cov}}$ yields what we will call the magnetostatic equation

\begin{equation}
\eb_3 + \iotaslash \eb_2 = G \eb_1\times\eb_2.\label{the-MS-eqn}
\end{equation}
From this, a ``local'' magnetostatic constraint \citep{skovoroda-2} (involving only derivatives within the surface) can be constructed:

\begin{equation}
g_{23} + \iotaslash g_{22} = 0,\label{local-MS-eqn}
\end{equation}
where we define $g_{ij} \equiv \eb_i\cdot\eb_j$.  Alternatively, this follows directly from $(\bnabla \psi \times \B_{\mathrm{cov}})\cdot \B_{\mathrm{con}} = 0$.  A noteworthy consequence of this constraint is discussed in Appendix \ref{appx-local-mhd}.
\subsection{The condition of quasi-axisymmetry (QAS)}

Quasi-symmetry can be stated as the condition that the magnetic field strength is a function of only a single helicity of the Boozer angles \citep{nuehrenberg-zille}, \ie $B = B(\psi, M\theta - N\varphi)$ where $M$ and $N$ are integers.  QAS is the $N = 0$ case, \ie simply the condition that the magnetic field strength is independent of the Boozer angle $\varphi$, \ie $\partial B/\partial \varphi = 0$.  Defining $E = G^2/B^2$ and using Eqn.~\ref{B-local-eqn}, we can express the quantity $E$ in terms of the surface metrics:

\begin{equation}
E = g_{33} + 2\iotaslash g_{23} + \iotaslash^2 g_{22}.\label{QAS-cond}
\end{equation} 

\noindent We can then restate QAS as $\partial E/\partial \varphi = 0$, \ie

\begin{equation}
\frac{\partial}{\partial \varphi}\left( g_{33} + 2\iotaslash g_{23} + \iotaslash^2 g_{22} \right) = 0.
\end{equation}

\subsection{The expansion about axisymmetry}

We write the coordinate mapping ${\boldsymbol x}(\psi, \theta, \varphi)$ as a series expansion in the small parameter $\epsilon$, 

\begin{equation}
{\boldsymbol x} = {\boldsymbol x}_0 + \epsilon {\boldsymbol x}_1 + \epsilon^2 {\boldsymbol x}_2 + \dots,
\end{equation}

\noindent where ${\boldsymbol x}_0$ corresponds to the zeroth-order axisymmetric magnetic field.  For simplicity, we will consider the current $G$ as fixed to its zeroth order values (there is no loss of generality as these functions can always be adjusted at zeroth order).  We however allow the deformation to modify $\iotaslash$, since we would like to determine how much ``external'' rotational transform can be achieved by the 3D shaping.

\begin{eqnarray}
\iotaslash(\psi) = \epsilon \iotaslash_1 + \epsilon^2 \iotaslash_2 + \dots,
\end{eqnarray}

\noindent where we have used the fact that a vacuum axisymmetric magnetic field has no rotational transform $\iotaslash_0 = 0$.  We also introduce the following notation for expanding the metrics:

\begin{equation}
g_{ij} = g_{ij}^{(0)} + \epsilon g_{ij}^{(1)} + \epsilon^2 g_{ij}^{(2)} + \dots.
\end{equation}

\subsubsection{${\cal O}(\epsilon^0)$}

The coordinate mapping at zeroth order is written

\begin{equation}
{\boldsymbol x}_0 = \Rh R_0(\theta, \psi) + \zh Z_0(\theta, \psi).\label{x0-eqn}
\end{equation}
Note that the unit vectors $\phih$, $\Rh$ and $\zh$ are defined with respect to the Boozer angle $\varphi$, (this choice will simplify the vector algebra at higher order)
\begin{eqnarray}
\Rh = \hat{\boldsymbol x} \cos(\varphi) + \hat{\boldsymbol y} \sin(\varphi),\label{Rh-eqn}\\
\phih = \hat{\boldsymbol y} \cos(\varphi) - \hat{\boldsymbol x} \sin(\varphi).\label{phih-eqn}
\end{eqnarray}
We emphasize that these are not the usual cylindrical basis vectors, as the Boozer toroidal angle only corresponds with the geometric toroidal angle at zeroth order; see discussion at the end of Appendix \ref{grad-shafranov-sec}.  Substituting Eqn.~\ref{x0-eqn} into Eqn.~\ref{the-MS-eqn}, and projecting along the $\phih$, $\Rh$ and $\zh$ directions, yields only a single nontrivial constraint
\begin{equation}
G\poiss{Z_0}{R_0} = R_0,\label{R0Z0-eqn}
\end{equation}
where we define 

\begin{equation}
\poiss{A}{B} \equiv \frac{\partial A}{\partial \psi}\frac{\partial B}{\partial \theta} - \frac{\partial B}{\partial \psi}\frac{\partial A}{\partial \theta}
\end{equation}
Of course, QAS is satisfied at this order, $\partial E_0/\partial \varphi = 0$, as

\begin{equation}
E_0 =  R_0^2.
\end{equation}
Note that Eqn.~\ref{R0Z0-eqn} does not constrain the choice of the zeroth order magnetic field very much.  In fact, we may freely choose not only the shape of the outer flux surface, but the shape of all the interior surfaces as well; see Appendix \ref{R0Z0-constraint-appx}.  If the reader is bothered by the fact that such surfaces are not proper flux surfaces, in the sense that, because $\iotaslash_0 = 0$, they are not unique, we remark that such non-uniqueness does not cause any mathematical inconsistencies at this order (the the contravariant form of ${\bf B}$ still correctly describes the field), and the uniqueness property will indeed by attained at higher order where the rotational transform is non-zero.

\subsubsection{${\cal O}(\epsilon^1)$}
So as to preserve the unit vectors, as mentioned above, we do not perturb the geometric angle $\phi$.  Instead, we include a (third) component of ${\boldsymbol x}_1$ in the  $\phih$ direction, \ie we write 
\begin{equation}
{\boldsymbol x}_1 = \Rh R_1(\theta, \psi, \varphi) + \zh Z_1(\theta, \psi, \varphi) + \phih \Phi_1(\theta, \psi, \varphi).\label{x1-eqn}
\end{equation}
As $\varphi$ is an ignorable coordinate in the first order magnetostatic equations, we will assume

\begin{eqnarray}
R_1 = \bar{R}_1(\theta, \psi) + 2\Re[\hat{R}_1(\theta, \psi) \exp(i N \varphi)],\\
Z_1 = \bar{Z}_1(\theta, \psi) + 2\Re[\hat{Z}_1(\theta, \psi) \exp(i N \varphi)],\\
\Phi_1 = \bar{\Phi}_1(\theta, \psi) + 2\Re[\hat{\Phi}_1(\theta, \psi) \exp(i N \varphi)],
\end{eqnarray}
where $N \neq 0$ is an integer, which will be interpreted as the field period number, since, as we will see, higher order corrections will only involve harmonics of this mode number.  Note that this is not the only way we can construct an $N$-period stellarator, as additional harmonics ($2N$, $3N$, \etc) could be included already at this order, but we make the above choice for simplicity.  The $\varphi$-averaged part of the local magnetostatic constraint (Eqn.~\ref{local-MS-eqn}, $\iotaslash_1 g_{22}^{(0)} + g_{23}^{(1)} = 0$) yields

\begin{equation}
\frac{\partial}{\partial \theta}\left(\frac{\bar{\Phi}_1}{R_0}\right) = -\frac{\iotaslash_1}{R_0^2}\left[ \left(\frac{\partial Z_0}{\partial \theta}\right)^2 + \left(\frac{\partial R_0}{\partial \theta}\right)^2 \right].\label{Phi-bar-1-eqn}
\end{equation}
Periodicity of $\bar{\Phi}_1/R_0$ yields the solubility constraint

\begin{equation}
\iotaslash_1 \oint \frac{d\theta}{R_0^2}\left[ \left(\frac{\partial Z_0}{\partial \theta}\right)^2 + \left(\frac{\partial R_0}{\partial \theta}\right)^2 \right] = 0,
\end{equation}
implying $\iotaslash_1 = 0$.  Then, from Eqn.~\ref{Phi-bar-1-eqn}, we find $\bar{\Phi}_1 = C_1 R_0$, where $C_1(\psi)$ is an arbitrary constant.  The only remaining useful information that can be obtained from the $\varphi$-average of Eqn.~\ref{the-MS-eqn} comes from its $\phih$ component, which gives a relationship between $\bar{R}_1$ and  $\bar{Z}_1$:

\begin{equation}
\bar{R}_1 = G\left( \poiss{\bar{Z}_1}{R_0}+ \poiss{Z_0}{\bar{R}_1} \right).\label{Phi1-bar-eqn}
\end{equation}
We are thus free to set $\bar{\Phi}_1 = \bar{R}_1 = \bar{Z}_1 = 0$ at this point; note the axisymmetric part of the first order solution could simply be absorbed into the zeroth order solution, and therefore does not represent interesting freedom in the QAS solution.  Furthermore, these components do not affect what follows, so there is no loss of generality. For the $\varphi$-dependent part of the first-order solution, we take components of the magnetostatic equation (Eqn.~\ref{the-MS-eqn}) in the $\phih$, $\Rh$ and $\zh$ directions to obtain

\begin{eqnarray}
iN \hat{\Phi}_1 + \hat{R}_1 = G\left( \poiss{\hat{Z}_1}{R_0}+ \poiss{Z_0}{\hat{R}_1} \right),\label{Phi1-eqn}\\
iN \hat{R}_1 - \hat{\Phi}_1 = G \poiss{\hat{\Phi}_1}{Z_0},\label{R1-eqn}\\
iN \hat{Z}_1 = G \poiss{R_0}{\hat{\Phi}_1}.\label{Z1-eqn}
\end{eqnarray}
A single second order equation for $\hat{\Phi}_1$ can be obtained from this system by substituting Eqns.~\ref{R1-eqn} and \ref{Z1-eqn} into Eqn.~\ref{Phi1-eqn}.  

\begin{equation}
(N^2 -1)\hat{\Phi}_1 + 2 G \poiss{Z_0}{\hat{\Phi}_1} = G^2\left (\poiss{R_0}{\poiss{R_0}{\hat{\Phi}_1}} + \poiss{Z_0}{\poiss{Z_0}{\hat{\Phi}_1}} \right).\label{P-pre-eqn}
\end{equation}
We are further able to simplify the resulting equation significantly by changing coordinates from ($\psi$, $\theta$) to ($R_0$, $Z_0$).  Defining

\begin{equation}
P_1(R_0, Z_0) = \hat{\Phi}_1(\psi(R_0, Z_0), \theta(R_0, Z_0)),
\end{equation}
and using Eqn.~\ref{R0Z0-eqn} (see also Appx.~\ref{R0Z0-coords-appx}), we obtain finally the simple equation

\begin{equation}
(N^2 - 1) P_1 = R_0^2 \Delta_0^* P_1,\label{P-eqn}
\end{equation}
where we encounter the familiar Grad-Shafranov operator

\begin{equation}
\Delta_0^* = R_0\frac{\partial}{\partial R_0}\left(\frac{1}{R_0}\frac{\partial}{\partial R_0}\right) + \frac{\partial^2}{\partial Z_0^2}.
\end{equation}
The QAS condition is enforced by setting $\partial E_1/\partial \varphi = 0$.  Using $E_1 =  g_{33}^{(1)}$ we obtain

\begin{equation}
iN R_0 (iN \hat{\Phi}_1 + \hat{R}_1 )= 0,\label{QAS-Phi-R}\\
\end{equation}
\ie $iN \hat{\Phi}_1 + \hat{R}_1 = 0$.  Using Eqn.~\ref{R1-eqn} we obtain from this a condition on $P_1$,

\begin{equation}
(N^2 - 1)P_1 + R_0 \frac{\partial P_1}{\partial R_0} = 0.\label{QAS-P}
\end{equation}
It can be easily verified that (except for a trivial $N = 1$ case corresponding to rotation; see Appendix \ref{no-global-QAS}) the only way to satisfy this condition across the entire plasma volume is to have $P_1 = 0$.  Global QAS is thereby proved impossible for vacuum fields, supporting the conclusions of \citet{garren-boozer-2}.  If, however, we require Eqn.~\ref{QAS-P} to be satisfied on a single surface, denoted $\psi = \psi_{bc}$, then Eqn.~\ref{QAS-P} constitutes an ``oblique'' (or more specifically, ``tangential-oblique'') boundary condition for the elliptic homogeneous second order equation \ref{P-eqn}.  This problem was first studied by Poincar\'{e} \citep{poincare}, and is therefore sometimes called ``Poincar\'{e}'s problem''.  The chief difficulty in the problem is due to the fact that the direction of the derivative is tangential to the boundary surface $\psi_{bc}$ at a discrete set of points.  In the present case, because Eqn.~\ref{P-eqn} is homogeneous, and the direction of the derivative does not rotate as the boundary curve is traced, it is known that exactly 2 linearly independent solutions exist (\cite{vekua}; see also Chapter III, Section 23 of \cite{miranda}), aside from the trivial null solution $P_1 = 0$.

For completeness we state the equations for $\hat{R}_1$ and $\hat{Z}_1$ in terms of $P_1$.  From Eqns.~\ref{R1-eqn}-\ref{Z1-eqn} we have

\begin{eqnarray}
iN \hat{R}_1 = P_1 -  R_0 \frac{\partial P_1}{\partial R_0},\label{R1-eqn-2}\\
iN \hat{Z}_1 = -R_0 \frac{\partial P_1}{\partial Z_0} .\label{Z1-eqn-2}
\end{eqnarray}

\subsubsection{${\cal O}(\epsilon^2)$}

We note that, at second order, the finite-toroidal-number components of the magnetostatic equation will generally be non-zero for $n = \pm 2N$ due to source terms that are quadratic in first order quantities.  We are free to choose the other components to be zero, and write the coordinate mapping in the same form as at first order, \ie ${\boldsymbol x}_2 = \Rh R_2(\theta, \psi, \varphi) + \zh Z_2(\theta, \psi, \varphi) + \phih \Phi_2(\theta, \psi, \varphi)$, with

\begin{eqnarray}
R_2 = \bar{R}_2(\theta, \psi) + 2\Re[\hat{R}_2(\theta, \psi) \exp(2 i N \varphi)],\\
Z_2 = \bar{Z}_2(\theta, \psi) + 2\Re[\hat{Z}_2(\theta, \psi) \exp(2 i N \varphi)],\\
\Phi_2 = \bar{\Phi}_2(\theta, \psi) + 2\Re[\hat{\Phi}_2(\theta, \psi) \exp(2 i N \varphi)].
\end{eqnarray}
To find $\iotaslash_2$ we can use the local magnetostatic constraint (Eqn.~\ref{local-MS-eqn}) at second order, \ie $\iotaslash_2 g_{22}^{(0)} + g_{23}^{(2)} = 0$.  Taking the $\varphi$-average, we obtain

\begin{multline}
\iotaslash_2\left[ \left(\frac{\partial Z_0}{\partial \theta}\right)^2 + \left(\frac{\partial R_0}{\partial \theta}\right)^2 \right] + R_0^2\frac{\partial}{\partial \theta}\left(\frac{\bar{\Phi}_2}{R_0}\right)\\
 + 2 \Re\left[ \frac{\partial \hat{R}_1^*}{\partial \theta}(i N\hat{R}_1 - \hat{\Phi}_1) +  i N \frac{\partial \hat{Z}_1^*}{\partial \theta} Z_1 +  \frac{\partial \hat{\Phi}_1^*}{\partial \theta} (\hat{R}_1 + i N \hat{\Phi}_1)\right] = 0.
\end{multline}
Integrating over $\theta$ (at constant $\psi$) we obtain $\iotaslash_2$ as a solubility constraint for $\bar{\Phi}_2$:

\begin{equation}
\iotaslash_2 = -\frac{\oint d\theta \frac{2}{R_0^2}\Re\left[ \frac{\partial \hat{R}_1^*}{\partial \theta}(i N\hat{R}_1 - \hat{\Phi}_1) +  i N \frac{\partial \hat{Z}_1^*}{\partial \theta} Z_1 +  \frac{\partial \hat{\Phi}_1^*}{\partial \theta} (\hat{R}_1 + i N \hat{\Phi}_1)\right]  }{\oint \frac{d\theta}{R_0^2}\left[ \left(\frac{\partial Z_0}{\partial \theta}\right)^2 + \left(\frac{\partial R_0}{\partial \theta}\right)^2 \right]}
\end{equation}

Following essentially the same procedure used at first order, we can obtain a single elliptic partial differential equation for $P_2(R_0, Z_0) = \hat{\Phi}_2(\psi(R_0, Z_0), \theta(R_0, Z_0))$,

\begin{multline}
(4N^2 - 1) P_2 - R_0^2 \Delta_0^* P_2\\
= 2 i N R_0 \poissRZo{\hat{Z}_1}{\hat{R}_1}  + R_0^2\left(\frac{\partial}{\partial Z_0} \poissRZo{\hat{R}_1}{\hat{\Phi}_1} + \frac{\partial}{\partial R_0}\poissRZo{\hat{\Phi}_1}{\hat{Z}_1} \right) ,\label{P2-eqn}
\end{multline}
where $\poissRZo{A}{B} = \partial_{R_0} A \partial_{Z_0} B - \partial_{R_0} B \partial_{Z_0} A$.  The condition of QAS is stated as

\begin{equation}
(2N^2 - 1)P_2 + R_0 \frac{\partial P_2}{\partial R_0} = -R_0 \poissRZo{\hat{\Phi}_1}{\hat{Z}_1} + \frac{iN}{R_0}\left[ (iN\hat{R}_1 - \hat{\Phi}_1)^2 + (iN\hat{Z}_1)^2\right].\label{QAS-P2}
\end{equation}
which is again an oblique boundary condition.  Eqn.~\ref{P2-eqn} with boundary condition \ref{QAS-P2} is solvable \citep{vekua, miranda}.

\subsubsection{${\cal O}(\epsilon^n)$, $n > 2$}

At higher order, there will be more equations to solve due to the nonlinear interaction of lower order contributions.  For instance, at third order, there will be nonzero $n = -3N, -N, 0, N, 3N$ components due to the source terms arising from the product of first and second order quantities.  However, these problems will be of the type as found at second order, and therefore must have solutions.  We conclude that QAS (at a single surface) can be satisfied to all orders in the expansion.

\section{Numerical solution of quasi-axisymmetric magnetic fields at first order}\label{numeric-sec}

We turn now to the numerical solution of Eqn.~\ref{P-eqn}, imposing the oblique boundary condition, Eqn.~\ref{QAS-P}, corresponding to the condition of QAS.  Note that, upon solving for $P_1$, the functions $\hat{R}_1$ and $\hat{Z}_1$ can then be immediately calculated from Eqns.~\ref{R1-eqn-2}-\ref{Z1-eqn-2} and, fixing $\epsilon$, the total mapping can be constructed as 

\begin{equation}
{\boldsymbol x} \approx \Rh (R_0 + 2 \epsilon \Re[\hat{R}_1 \exp(i N \varphi)] ) +  \zh (Z_0 + 2 \epsilon \Re[\hat{Z}_1 \exp(i N \varphi)]) + 2 \phih \epsilon  \Re[\hat{\Phi}_1 \exp(i N \varphi)].\label{x-approx}
\end{equation}
Due to the change of variables to ($R_0$, $Z_0$) space, the only knowledge of the zeroth order solution needed is the shape of the outer flux surface, which defines the computational domain.  If the first order deformation is specified, along with the amplitude $\epsilon$, then ${\boldsymbol x}$ is determined by Eqn.~\ref{x-approx}, in the coordinates $R_0$, $Z_0$ and $\varphi$.  We will henceforth drop the subscripts and work directly with the following equations for $P$ in the $R$-$Z$ plane:

\begin{eqnarray}
(N^2 - 1) P + 3 R \frac{\partial P}{\partial R} = (\hat{\bf R} \partial_R +  \zh \partial_Z) \cdot \left(\Rh R^2 \frac{\partial P}{\partial R} + \zh R^2 \frac{\partial P}{\partial Z}\right),\label{P-eqn-2}\\
(N^2 - 1)P + R \frac{\partial P}{\partial R} = 0.\label{QAS-P-2}
\end{eqnarray}
Note that Eqn.~\ref{P-eqn-2} is simply Eqn.~\ref{QAS-P}, with the diffusive term rewritten in conservative form.

The numerical method used to solve Eqn.~\ref{P-eqn-2} with boundary condition \ref{QAS-P-2} is based on a standard finite element method approach to a Neumann boundary value problem.  As usual, the function $P$ is expressed as a sum of basis functions that have finite support over mesh elements.  Multiplying Eqn.~\ref{P-eqn-2} by a basis function and integrating over the domain, one obtains a matrix equation containing the so-called stiffness matrix.  Part of the stiffness matrix is due to the flux at the boundary (arising from the conservative term on the right hand side of Eqn.~\ref{P-eqn-2}), and involves the normal derivative of $P$.  This flux term is evaluated using our boundary condition, \ref{QAS-P-2}.  That is, the derivative term in Eqn.~\ref{QAS-P-2} is rewritten in terms of components that are tangential and normal to be boundary:

\begin{equation}
(N^2 - 1) P + \gamma(l) \frac{\partial P}{\partial n} + \beta(l) \frac{\partial P}{\partial l}= 0.\label{QAS-P-3}
\end{equation}
The functions $\gamma$ and $\beta$ depend on the shape of the boundary.  Eqn.~\ref{QAS-P-3} is divided by $\gamma$ to obtain an expression for the normal derivative, and the resulting equation is discretized over a one-dimensional mesh whose nodes match the boundary nodes of the finite element mesh.  The result is used to evaluate the relevant contribution the stiffness matrix obtained from the discretization of Eqn.~\ref{P-eqn-2}.

The only difficulty in this procedure arises from the fact that the coefficient of the normal derivative, $\gamma(l)$, passes through zero at a discrete set of boundary points (those where the normal vector points in the $\hat{\bf z}$ direction) and so the flux is not determined by Eqn.~\ref{QAS-P-3} at those points.  To overcome this, we only need to avoid these points when choosing the nodes of the mesh that lie on the boundary.

The final result of the above procedure is a homogeneous matrix equation.  The two eigenfunctions with eigenvalues that tend toward zero (with increasing resolution) are identified as the solutions.  The solutions can (due to their linear independence) be chosen such that one is even and the other is odd in a suitably defined geometric angle $\theta$.  A sum of the two solutions can be taken, and the relative phase and amplitude can be adjusted to maximize the rotational transform.

\begin{figure}
\includegraphics[width=0.29\columnwidth]{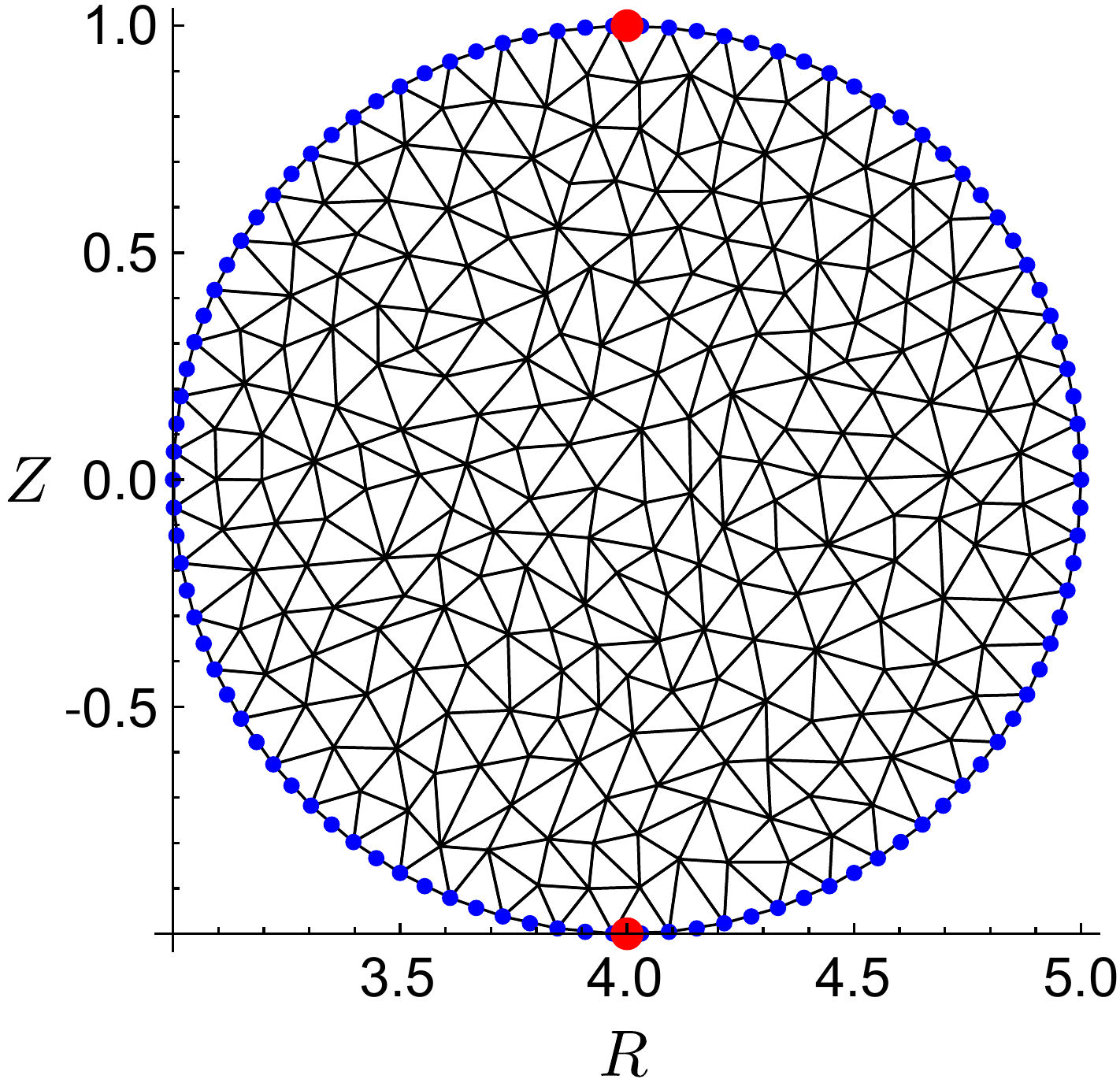}
\includegraphics[width=0.34\columnwidth]{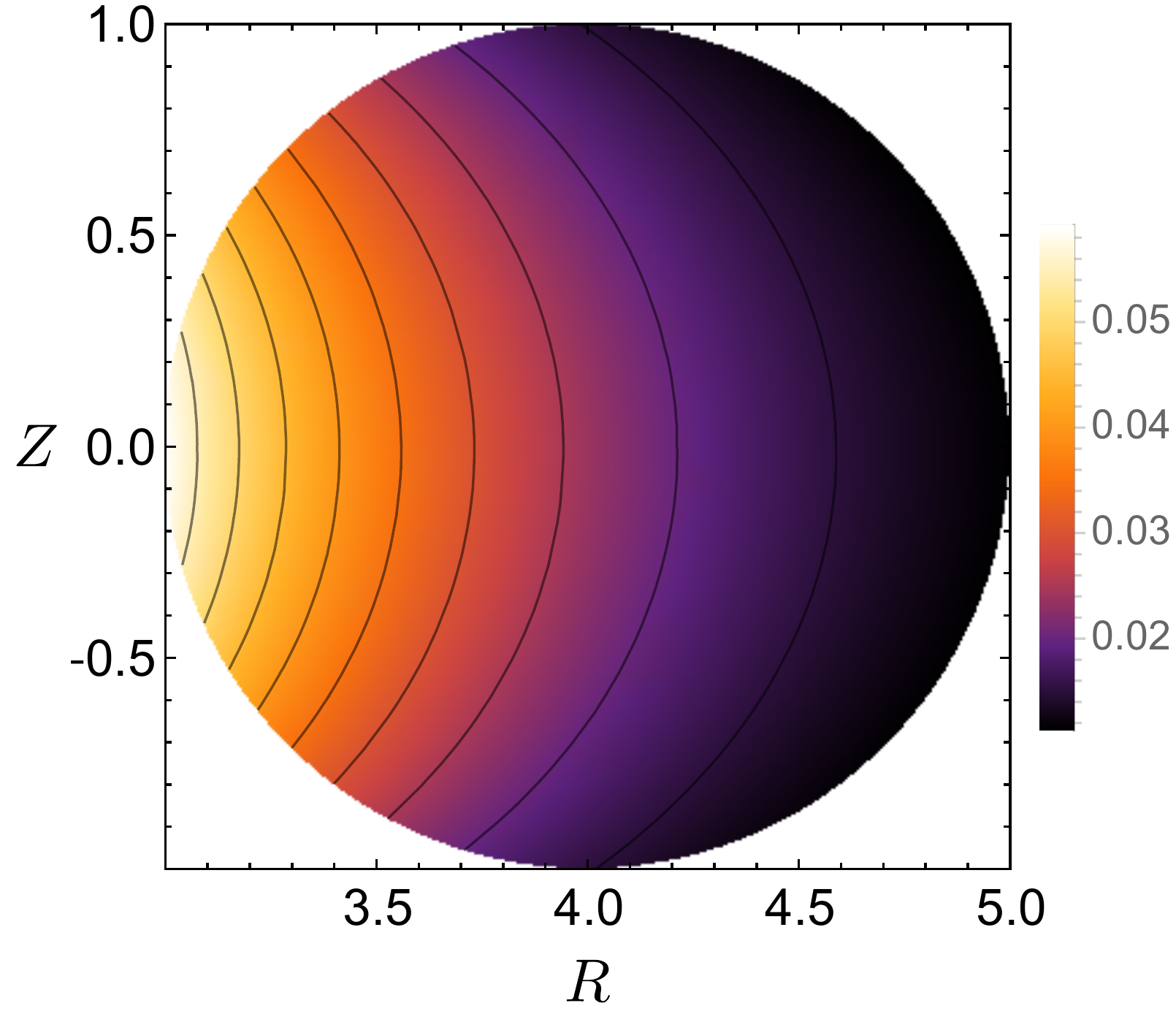}
\includegraphics[width=0.34\columnwidth]{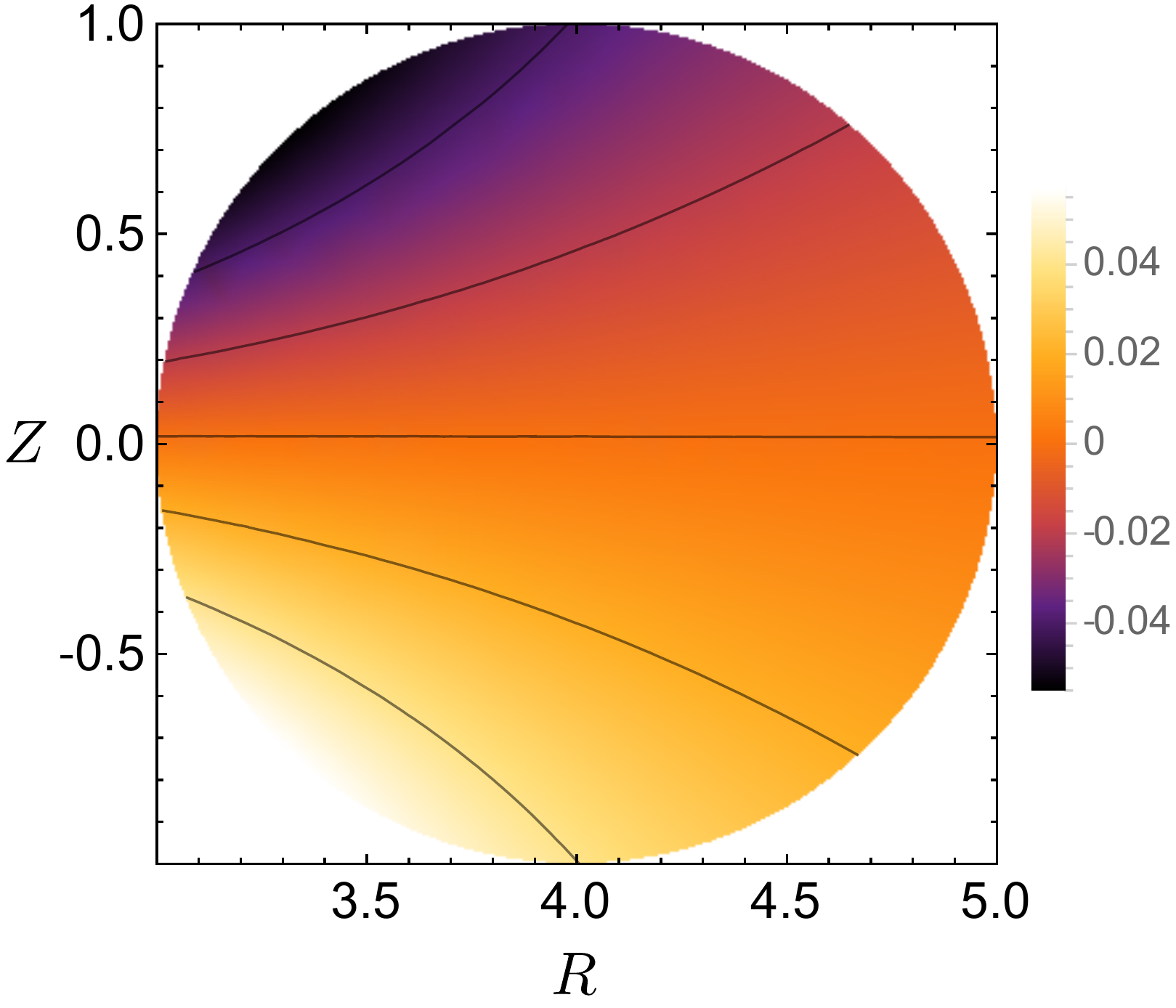}
\caption{Solution for circular boundary with $N=2$ chosen.  On the left is a visualization of a circular second-order mesh, with boundary nodes indicated with blue dots, and the two points of tangency indicated with red dots.  The numerical solution, including mesh generation, is done with the recently introduced finite-element-method capabilities of Mathematica 10.  In the center and right are two independent solutions found for $P(R_0, Z_0)$.  Note that due to the up-down mirror symmetry of the solution, the two solutions can be chosen as even (center) and odd (right).}
\label{mesh-figure}
\end{figure}

Let us illustrate the above procedure with the simple example of a circular boundary centered at $R = R_c$ with radius $1$.  This case yields $\gamma(l) = [R_c + \cos(l)]\cos(l)$, and $\beta(l) = -[R_c + \cos(l)]\sin(l)$.  The resulting mesh and two independent solutions are depicted in Fig.~\ref{mesh-figure}.

\begin{figure}
\begin{center}
\includegraphics[width=0.65\columnwidth]{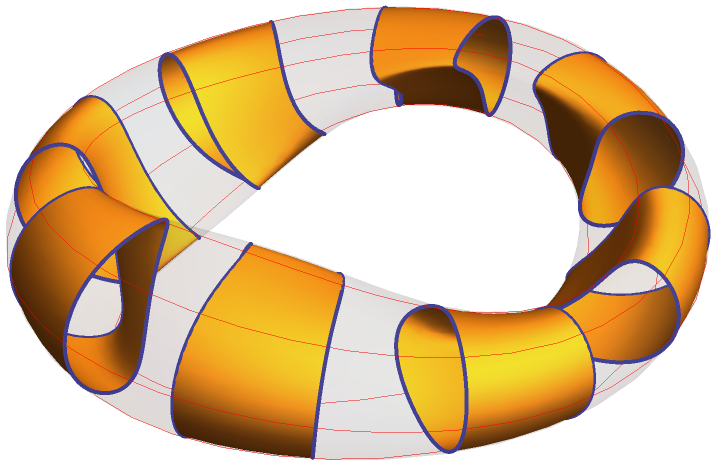}
\end{center}
\caption{Visualization of boundary surface shape up to first order, assuming circular zeroth-order shape (same case as Fig.~\ref{mesh-figure}).  Here $N=2$ and $\epsilon = 2.0$; note that the eigenfunctions are normalized to small numerical absolute value, as shown in Fig.~\ref{mesh-figure}, so the total deformation is small even though $\epsilon$ is not.  The toroidal cuts and lines of constant poloidal and toroidal angle are included for purely stylistic reasons.}
\label{surface-viz-fig}
\end{figure}

The even and odd solutions are superposed to create a total solution $P = A P_e + B P_o$, where $A$ and $B$ are arbitrary complex constants.  To obtain finite rotational transform for the present example, it is necessary to assume a non-zero phase shift between these amplitudes.  Taking $A= 1$ and $B = \exp(-i\pi/2)$ and fixing $\epsilon$ the surface shape, according to Eqn.~\ref{x-approx}, is plotted in Fig.~\ref{surface-viz-fig}.

\begin{figure}
\includegraphics[width=0.45\columnwidth]{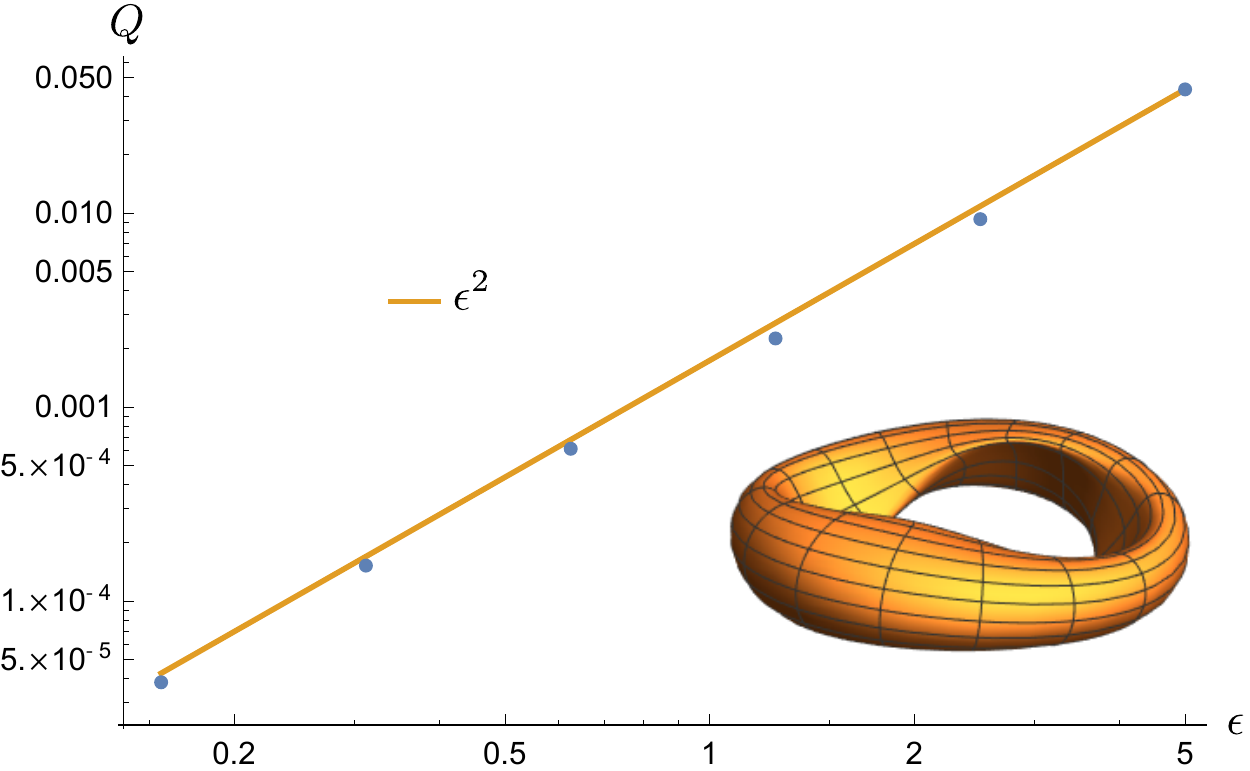}
\includegraphics[width=0.45\columnwidth]{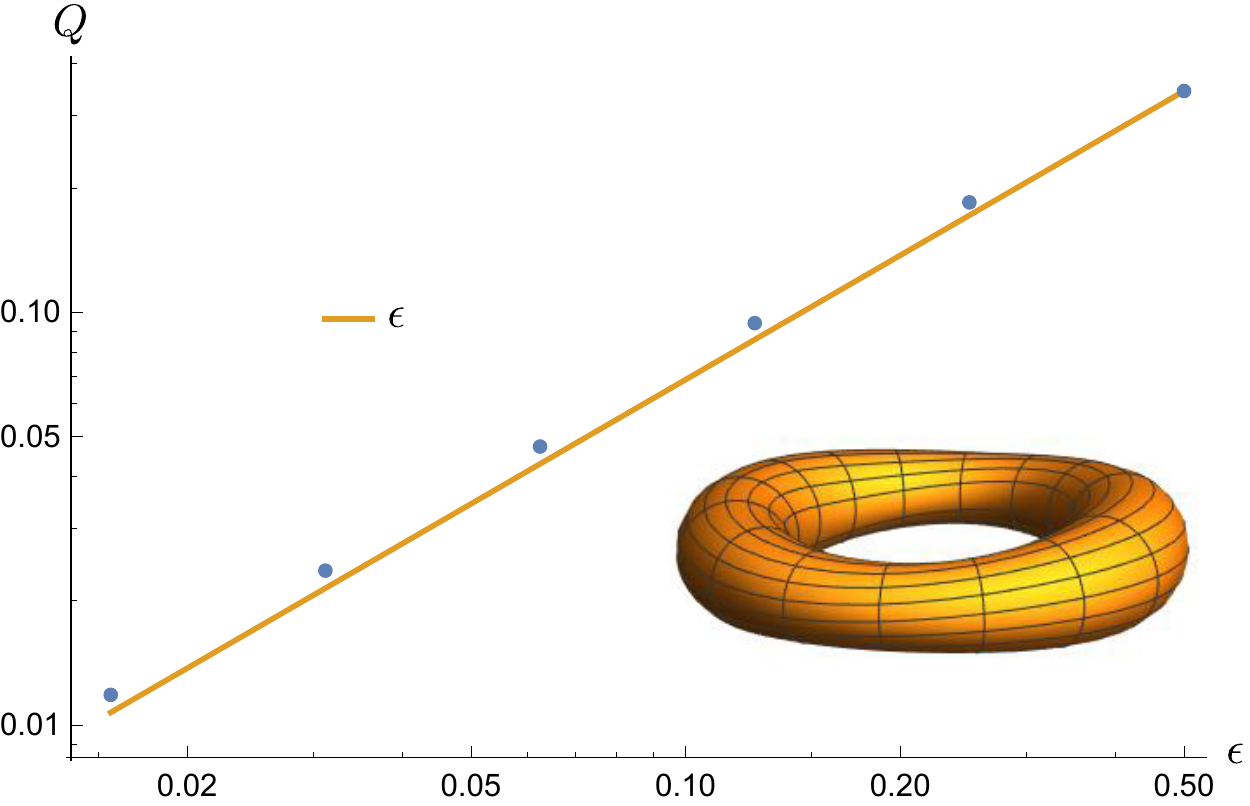}
\caption{Demonstration that QAS is satisfied to first order in size of non-axisymmetric perturbation; the measure of deviation $Q$, defined by Eqn.~\ref{Q-eqn}, is computed only at the surface where the QAS condition is applied.  Here the field period number is $N = 2$ and the zeroth order flux surface shape is circular with an aspect ratio of $4$.  Solution of Eqn.~\ref{P-eqn-2} with boundary condition Eqn.~\ref{QAS-P-2} leads to expected scaling with $\epsilon^2$, as shown in the first plot.  In the second plot, a ``control'' (non-QAS) deformation is used for comparison ($\Phi_1 = 0$, $R_1 = 0.4 \cos(\vartheta - 2\varphi)$ and $Z_1 = 0.4 \sin(\vartheta - 2\varphi)$, where $\vartheta$ is a geometric poloidal angle), and it is found that QAS is broken at first order in $\epsilon$, as expected.  Visualization of surface shape (for largest value of $\epsilon$) is included within each plot.  (Note that $\epsilon$ is somewhat arbitrary as the solutions are linear and subject to arbitrary normalization.)}
\label{QAS-confirm-figure}
\end{figure}

To independently confirm the result, the surface shape, determined by Eqn.~\ref{x-approx}, can be used as input for the VMEC code.  This shape specification has a free parameter $\epsilon$, which can be varied to test the degree to which the result satisfies the QAS condition.  Since the solution is only correct to first order, the ``exact'' solution obtained from VMEC will be expected to deviate from Quasiaxisymmetry at second order, on the appropriate flux surface.  Therefore, we conduct a series of runs with increasing strength of perturbation, $\epsilon$.  The resulting solution is then passed to a separate code ``BOOZ\_XFORM'' \citep{booz-xform}, which computes the components of $|{\bf B}|$ in Boozer angles separately on each flux surface.  Examining these components for the outer surface, we find that the ratio of the non-QAS part to the full field, does indeed scale as $\epsilon^2$.  For comparison, the scaling of a control (non-QAS, non-axisymmetric) deformation is also computed, and found to be linear in $\epsilon$ as expected; See Fig.~\ref{QAS-confirm-figure}.  In this figure the quantity $Q$ measures the deviation from QAS, and is defined in terms of the Fourier component of the magnetic field magnitude expressed in Boozer coordinates $\hat{B}_{mn}$, where $m$ and $n$ are the poloidal and toroidal mode numbers respectively:

\begin{equation}
Q = \frac{\left(\sum_{m, n\neq 0}|\hat{B}_{mn}|^2\right)^{1/2}}{\left(\sum_{m, n}|\hat{B}_{mn}|^2\right)^{1/2}}.\label{Q-eqn}
\end{equation}

\section{Omnigeneity}\label{gen-qas-sec}

Omnigeneity \citep{hall-macnamara-omnig} is the more general condition of having zero average radial particle drift.  This can be shown to be equivalent to the geometric condition that points of equal magnetic field strength have constant separation in Boozer angles, \ie the separation ($\Delta \theta$, $\Delta \varphi$) is independent of the field line chosen (see also \cite{landreman} for further discussion of this).  It has been argued that exactly omnigenous configurations, must be quasi-symmetric if they are to be analytic \citep{cary-shasharina}.  The same authors point out, however, that analytic fields can be specified that are arbitrarily close to being omnigenous, while also being far from quasi-symmetric.  One can conclude that omnigeneity is a useful target for optimization.  Indeed, it is fair to say, roughly speaking, that nearly quasi-symmetric configurations represent a small subspace of nearly omnigenous configurations.  Therefore, it may be useful to be able to impose omnigeneity instead of quasi-symmetry as a boundary condition in our expansion near axi-symmetry.  As will be shown, we can impose an arbitrary modification to the magnetic field strength, $E_1(\theta, \varphi)$.  This function can, in particular, be chosen such that $E_0 + \epsilon E_1$ satisfies the condition of omnigeneity.  We write $E_1$ as a Fourier series,
\begin{equation}
E_1 = \bar{E}_1(\psi, \theta) + \sum_{n\neq0} \hat{E}_1(\psi, \theta, n) \exp(i n \varphi),
\end{equation}
and first consider the $\varphi$-independent part, $\bar{E}_1$.  Note that the magnetostatic equation, Eqn.~\ref{the-MS-eqn}, was already solved for the QAS problem, and yields $\bar{\Phi}_1 = C_1 R_0$ and $\iotaslash_1 = 0$.  From Eqn.~\ref{QAS-cond}, we then find that $\bar{E}_1 = 2 R_0 \bar{R}_1$, \ie

\begin{equation}
\bar{R}_1 = \frac{\bar{E}_1}{2 R_0}.
\end{equation}
We now can rewrite Eqn.~\ref{Phi1-bar-eqn}, which relates $\bar{Z}_1$ to $\bar{R}_1$, using Eqn.~\ref{R0Z0-eqn}, in terms of independent variables $Z_0$ and $R_0$ as follows

\begin{equation}
\frac{\partial \bar{Z}_1}{\partial Z_0} = \bar{R}_1 - \frac{\partial \bar{R}_1}{\partial R_0}.\label{Phi1-bar-eqn-2}
\end{equation}
Thus $\bar{Z}_1$ is determined by $\bar{R}_1$ via Eqn.~\ref{Phi1-bar-eqn-2}, completing the $\varphi$-independent part of the solution.  For the $\varphi$-dependent part of the solution, we essentially only need to add a non-zero contribution to the right-hand-side of Eqn.~\ref{QAS-P}, noting, however, that the resulting equation must be satisfied for all toroidal mode numbers.  We therefore modify the notation $N \rightarrow n$, and obtain the following PDE for $P_1(R_0, Z_0, n)$, with $n\neq 0$:

\begin{equation}
(n^2 - 1) P_1 = R_0^2 \Delta_0^* P_1.\label{P-eqn-general}
\end{equation}
The boundary condition filling the role of Eqn.~\ref{QAS-P}, now with an inhomogeneity proportional to $\hat{E}_1(\psi, \theta, n)$, becomes (for $n \neq 0$):

\begin{equation}
R_0 (n^2-1) P_1 + R_0^2 \frac{\partial P_1}{\partial R_0}= -i n \hat{E}_1.\label{P-bc-general}
\end{equation}
We note first, that the existence of at least one solution is guaranteed for this non-homogeneous oblique derivative problem (\cite{vekua}; see also Chapter III, Section 23 of \cite{miranda}).  This problem is more difficult to solve than the QAS problem, since the equations for all necessary Fourier components of $E_1$ must be solved.  Furthermore, it seems necessary to solve for the $\theta(R_0, Z_0)$ to be able to construct a function $E_1$ such that the total field is omnigenous; this was not necessary for the QAS problem since only the shape of the zeroth-order boundary flux surface was needed to solve the first order system.  

We may draw an interesting conclusion at this point, namely that the QAS problems, which can be indexed by $n \neq 0$, are valid homogeneous solutions of the omnigenous problem, \ie the corresponding deformations do not modify the magnitude of the magnetic field (by construction), so they can be added to a solution of the general first order problem (Eqns.~\ref{P-eqn-general} and \ref{P-bc-general}), and the result will still satisfy Eqns.~\ref{P-eqn-general} across the volume, and Eqn.~\ref{P-bc-general} on the boundary.  The QAS deformations therefore represent an arbitrary freedom for omnigenous solutions.

\section{Conclusions}\label{conclusion-sec}

One of the perhaps surprising outcomes of this work is the abundance of QAS solutions.  Instead of being isolated, they form a continuum in the vicinity of axisymmetric solutions.  The parameter space for this continuum consists of the (1) the space of axi-symmetric flux-surface shape functions, (2) the discrete parameter $N$, defining the stellarator field period number, and (3) the two complex amplitudes of the solutions of the oblique-derivative problem.  From one perspective, the size of this space might be expected:  The axisymmetric part of our solution satisfies QAS in a trivial sense, and globally.  So much of the freedom in the solution corresponds to the part of the solution that is known to be exceptional.  If, as is true for the near-axis expansion of \citep{garren-boozer-2}, the results found in the vacuum case are also valid for the non-vacuum case, a striking and simple conclusion could be drawn -- namely, that a large space of QAS stellarator configurations are accessible via suitably small deformations of any tokamak.

We reach another interesting conclusion, following the discussion of Sec.~\ref{gen-qas-sec}, namely that an alternative problem specification is possible, by fixing the perturbation in magnetic field strength at the boundary flux surface, as a function of Boozer angles.  A solution is guaranteed to exist for this problem, and we note that an arbitrary QAS deformation may be added to this solution, such that the total solution still satisfies the field-strength boundary condition.  Furthermore, the field strength may be chosen such that a generalized form of QAS (omnigeneity) is satisfied, significantly broadening the space of ``good'' configurations that can be found in the neighborhood of axi-symmetry.  Although a deformation can only satisfy exact QAS at one surface, it remains an open question whether omnigeneity can be satisfied throughout the volume.

As a final note, it is worth remarking on how our perturbed solutions, which can be thought of as nearly 2D, relate to fully 3D magnetic fields, which are known to suffer from singularities associated with rational rotational transform that can break up flux surfaces.  Since our treatment begins with the assumption of flux surfaces, it is fair to ask to what extent this assumption could be broken in actual configurations unconstrained by this assumption.  However, we can argue that this will not be a problem, for at least sufficiently small deviations from axisymmetry.  This is because the obtained rotational transform is necessarily small, scaling as $\epsilon^2$, and so the rational surfaces must be very high order, \ie $\iotaslash = n/m$ with $m \gg 1$.  Such surfaces, as argued by \citet{rosenbluth-sagdeev-taylor}, give rise to island chains that have exponentially small width.

{\bf Acknowledgements.}  Thanks to Sophia Henneberg for helpful conversations, and Sam Lazerson and Joachim Geiger for assistance with the VMEC code.  This work has been carried out within the framework of the EUROfusion Consortium and has received funding from the Euratom research and training programme 2014-2018 under grant agreement No 633053.  The views and opinions expressed herein do not necessarily reflect those of the European Commission.

\appendix
\section{Magnetic geometry}\label{geometry-appx}

Given arbitrary toroidal coordinates ($\psi$, $\theta$, $\varphi$), whose handedness is $\bnabla\varphi\cdot(\bnabla\psi\times\bnabla\theta) > 0$, it is the convention to work with local basis vectors, ($\bnabla\psi$, $\bnabla\theta$, $\bnabla\varphi$) and ($\eb_1 = \partial {\boldsymbol x}/\partial \psi$, $\eb_2 = \partial {\boldsymbol x}/\partial \theta$, $\eb_3 = \partial {\boldsymbol x}/\partial \varphi$).  The metric components are defined in the usual way

\begin{equation}
g_{ij} \equiv \eb_i\cdot\eb_j,
\end{equation}

\noindent and the Jacobian for these coordinates is

\begin{equation}
\Jac = \frac{1}{\bnabla\psi\cdot(\bnabla\theta\times\bnabla\varphi)} = \eb_1\cdot (\eb_2\times\eb_3)
\end{equation}

\noindent Additionally, assigning $(u_1, u_2, u_3) = (\psi, \theta, \varphi)$, we see that $\eb_i\cdot\bnabla u_j = \delta_{ij}$, and the following identities are easily verified

\begin{eqnarray}
\eb_1 = \Jac(\bnabla u_2\times\bnabla u_3),\quad \eb_2 = \Jac(\bnabla u_3\times\bnabla u_1),\quad \eb_3 = \Jac(\bnabla u_1\times\bnabla u_2),\\
\bnabla u_1 = \frac{\eb_2\times\eb_3}{\Jac},\quad \bnabla u_2 = \frac{\eb_3\times\eb_1}{\Jac},\quad \bnabla u_3 = \frac{\eb_1\times\eb_2}{\Jac}.
\end{eqnarray}

\section{Useful forms of $\Jac$ and $B$}

Taking $B^2 = \B_{\mathrm{cov}} \cdot \B_{\mathrm{con}}$ gives

\begin{equation}
\Jac B^2 = G + \iotaslash I.\label{JacB-eqn-1}
\end{equation}

\noindent Taking $B^2 = \B_{\mathrm{con}} \cdot \B_{\mathrm{con}}$ gives

\begin{equation}
\Jac^2 B^2 = |\eb_3 + \iotaslash \eb_2|^2 = g_{33} + 2\iotaslash g_{23} + \iotaslash^2 g_{22}.\label{JacB-eqn-2}
\end{equation}

\noindent Using Eqn.~\ref{JacB-eqn-1}-\ref{JacB-eqn-2} we can express the magnetic field strength ``locally'' (in terms of only surface metrics),

\begin{equation}
\frac{(G+\iotaslash I)^2}{B^2} = g_{33} + 2 \iotaslash g_{23} + \iotaslash^2 g_{22}.\label{B-local-eqn}
\end{equation}

\noindent Using Eqn.~\ref{JacB-eqn-1}-\ref{JacB-eqn-2} we can also express the Jacobian locally,

\begin{equation}
\Jac (G+\iotaslash I) = g_{33} + 2 \iotaslash g_{23} + \iotaslash^2 g_{22}.\label{Jac-local-eqn}
\end{equation}

\section{General axisymmetric fields in Boozer coordinates}\label{axi-appx}

The condition of axi-symmetry can be stated as the condition that the $\hat{\boldsymbol R}$, $\hat{\bphi}$ and $\hat{\boldsymbol z}$ components of $\B$ are independent of $\phi$.  This implies that the flux surfaces must be themselves axisymmetric, so $\hat{\bphi}\cdot \bnabla \psi = 0$, \ie $\partial_\phi \psi = 0$.  Then from $\partial_\phi (\hat{\bphi}\cdot\B_{\mathrm{cov}} )= 0$ we obtain

\begin{equation}
G \frac{\partial ^2 \varphi}{\partial \phi^2} + I \frac{\partial^2 \theta}{\partial \phi^2} = 0.
\end{equation}  
Integrating, using $\varphi = \phi + \nu$, and  periodicity in $\phi$ we obtain

\begin{equation}
G\frac{\partial \nu}{\partial \phi} + I \frac{\partial \theta}{\partial \phi} = 0.\label{axi-eqn-1}
\end{equation}
Now, taking the $\partial_\phi (\hat{\boldsymbol R}\cdot B_{\mathrm{con}} )= 0$, we likewise obtain

\begin{equation}
\frac{\partial \theta}{\partial \phi} - \iotaslash \frac{\partial \nu}{\partial \phi} = 0.\label{axi-eqn-2}
\end{equation}
Note that Eqns.~\ref{axi-eqn-1}-\ref{axi-eqn-2} are linearly independent (as guaranteed by Eqn.~\ref{JacB-eqn-1}), so we can conclude that

\begin{equation}
\frac{\partial \theta}{\partial \phi} = \frac{\partial \nu}{\partial \phi} = 0.\label{axi-eqn-3}
\end{equation}

\subsection{Axisymmetric fields in the direct formulation}\label{grad-shafranov-sec}

Switching to the direct formulation, let us consider magnetic coordinates $\psi$, $\theta$, $\varphi$ as functions of cylindrical coordinates $R$, $Z$ and $\phi$.  Here $Z$ is the distance along the axis of symmetry (of the zeroth order solution), $R$ is the distance from the origin in the plane perpendicular to $Z$, and $\phi$ is the geometric toroidal angle measuring the rotation about the $Z$ axis.  Writing $\B_{\mathrm{con}} = \B_{\mathrm{cov}}$ we have

\begin{equation}
G(\psi) \bnabla\varphi + I(\psi) \bnabla\theta + K(\psi, \theta, \varphi) \bnabla \psi= \bnabla\psi \times \bnabla\theta - \iotaslash \bnabla \psi \times \bnabla \varphi,\label{MS-direct-eqn}
\end{equation}
In a vacuum axisymmetric field we have $\iotaslash = K = I = 0$ and $dG/d\psi = 0$, \ie

\begin{equation}
G\bnabla\varphi = \bnabla\psi \times \bnabla\theta,\label{MS-direct-eqn-vacuum}
\end{equation}

By axisymmetry, $\psi$ and $\theta$ are independent of $\phi$ (see Appendix \ref{axi-appx}), whereas $\varphi$ can generally be expressed as

\begin{equation}
\varphi = \phi + \nu(R, Z)
\end{equation}
The right hand side of Eqn.~\ref{MS-direct-eqn-vacuum} has only a toroidal component, so the $\Rh$ and $\zh$ components of the equation require that $\nu$ is a constant, which we set to zero.  The $\hat{\bphi}$ component of Eqn.~\ref{MS-direct-eqn-vacuum} yields the only nontrivial constraint.

\begin{equation}
\frac{G}{R} = \poissRZ{\theta}{\psi},
\end{equation}
where $\poissRZ{A}{B} = \partial_R A \partial_Z B - \partial_R B \partial_Z A$.  To summarize,

\begin{eqnarray}
\psi = \psi(R, Z),\\
\theta = \theta(R, Z),\\
\varphi = \phi.\label{phi-0-eqn}
\end{eqnarray}

\noindent  We note that Eqn.~\ref{phi-0-eqn} can be interpreted (within the context of the expansion about axisymmetry) to mean that the zeroth-order geometric toroidal angle is equal to the Boozer toroidal angle, \ie $\phi_0 = \varphi$; the geometric angle $\phi$ can be defined exactly in terms of the coordinate mapping via the relation $\tan(\phi) = (\hat{\bf y}\cdot {\boldsymbol x})/(\hat{\bf x}\cdot{\boldsymbol x})$.  One can observe that this geometric angle will acquire corrections at higher order, $\phi_1$, $\phi_2$, {\em etc.}, due to the modifications to the coordinate mapping.  However, we can avoid solving for these corrections by defining the unit vectors $\Rh$ and $\phih$ in terms of the Boozer angle $\varphi$ (Eqns.~\ref{Rh-eqn}-\ref{phih-eqn}) and allowing the perturbation of the coordinate mapping to have a component in the $\phih$ direction; see Eqn.~\ref{x1-eqn}.  We hope these conventions do not cause confusion, but they significantly simplify the vector algebra that follows, since it avoids unnecessary complications associated with expanding the conventionally defined unit vector, $\Rh(\phi) = \hat{\bf x} \cos(\phi) + \hat{\bf y} \sin(\phi)$.
\subsection{Freedom at zeroth order according to Eqn.~\ref{R0Z0-eqn}}\label{R0Z0-constraint-appx}

Consider a set of nested arbitrarily-shaped axisymmetric surfaces, labeled by the variable $\rho$; let these correspond to constant-$\psi$ surfaces. The function $\psi(\rho)$ is to be determined, as follows.  Let the variable $l$ be the arc length parameterizing a contour of constant $\rho$, such that $(\hat{\bf \phi} \times \bnabla \rho)\cdot \bnabla l > 0$.  Then Eqn.~\ref{R0Z0-eqn} becomes

\begin{equation}
\frac{d\psi}{d \rho}\frac{\partial \theta}{\partial l} = \frac{G}{R_0 |\bnabla \rho|} \equiv F(\rho, l)
\end{equation}
using $\oint \partial \theta/\partial l dl = -2 \pi$ we find

\begin{eqnarray}
\psi = \frac{1}{2\pi}\int_0^{\rho} d\rho^\prime \oint dl F(\rho^\prime, l),\\
\theta = \theta_0(\rho) -2\pi \int_0^{l} dl^\prime \frac{F(\rho, \ell^\prime)}{\oint dl^{\prime\prime} F(\rho, l^{\prime\prime})}.
\end{eqnarray}
where we use that the toroidal flux on axis $\psi(\rho = 0)$ is zero.  It is thus shown that the choice of an arbitrary set of nested axisymmetric surfaces is consistent with Eqn.~\ref{R0Z0-eqn}, and this choice determines the spatial dependence of Boozer coordinates $\psi$ and $\theta$ up to an arbitrary angular shift $\theta_0(\rho)$, which can be inverted to find $R_0$ and $Z_0$ as functions of $\psi$ and $\theta$.

\section{Using $R_0$ and $Z_0$ as coordinates}\label{R0Z0-coords-appx}

The differential operators of Eqn.~\ref{P-pre-eqn} take the form of Possion brackets involving the functions $R_0$ and $Z_0$.  Such operators can be interpreted as 2D advection along the $\bnabla R_0$ and $\bnabla Z_0$ directions, which suggests that it could be fruitful to reformulate the equations, using $R_0$ and $Z_0$ as the independent variables, especially given that these variables are well-behaved coordinates.  Furthermore, it will be useful, for the purposes of numerically solving the first-order system, to be able to re-use the domain defined already at zeroth order.

The following relations, found using Eqn.~\ref{R0Z0-eqn}, may make the derivation of Eqn.~\ref{P-eqn} more transparent.  For any function $f$, we have

\begin{eqnarray}
\poiss{Z_0}{f} = \frac{R_0}{G}\frac{\partial f}{\partial R_0},\\
\poiss{R_0}{f} = -\frac{R_0}{G}\frac{\partial f}{\partial Z_0}.
\end{eqnarray}

\section{Consequence of the local magnetostatic constraint on the shape of QAS configurations}\label{appx-local-mhd}

A differential constraint involving only in-surface derivatives can be derived directly from the condition $\B_{\mathrm{con}} \cdot ( \B_{\mathrm{cov}} \times \bnabla \psi ) = 0$.  This, which we called the ``local magnetostatic constraint'', is given in Eqn.~\ref{local-MS-eqn} and repeated here:

\begin{equation}
g_{23} + \iotaslash g_{22} = 0.\label{local-MS-eqn-2}
\end{equation}
At first order in the expansion about axi-symmetry, Eqn.~\ref{local-MS-eqn-2} yields for $N \neq 0$ 

\begin{equation}
R_0 \frac{\partial \hat{\Phi}_1}{\partial \theta} + \frac{\partial R_0}{\partial \theta}\left(iN \hat{R}_1 -\hat{\Phi}_1\right) + iN\frac{\partial Z_0}{\partial\theta}\hat{Z}_1 = 0.
\end{equation}
This equation, combined with the condition of QAS, Eqn.~\ref{QAS-Phi-R}, gives

\begin{equation}
\frac{\partial}{\partial \theta}\left(\hat{\Phi}_1 R_0^{N^2-1}\right) = -iN\frac{\partial Z_0}{\partial \theta} R_0^{N^2-1} \hat{Z}_1,
\end{equation}
which can be integrated over $\theta$ to give the solubility constraint

\begin{equation}
\oint d\theta \frac{\partial Z_0}{\partial \theta} R_0^{N^2-1} \hat{Z}_1 = 0.
\end{equation}
This constitutes a rather strong constraint on the behavior of the function $\hat{Z}_1(\theta)$ at low aspect ratio and large field period number $N$ since in that limit the factor $R_0^{N^2-1}$ varies strongly in magnitude.  This leads to the qualitative behavior around the surface where QAS is satisfied (indeed, observed in our numerical solutions) of the perturbation favoring the ``inboard'' side of the device (locations of small $R$), \ie having greater amplitude there; the surface shape on the outboard side, in such cases, remains mostly unperturbed from its axisymmetric shape.  At small $N$ (\ie $N=2$) and large aspect ratio (\ie when the variation in the overall magnitude of $R_0$ is small) this effect is less noticeable.
\section{Non-existence of global QAS magnetic fields}\label{no-global-QAS}

We show here that the QAS condition can be satisfied globally only for the special cases of $N = 1$, and, in that case, the deformation merely corresponds to trivial transformations that preserve axisymmetry.  We demonstrate this at first order in the expansion.  It is easy to integrate the QAS condition (it can be interpreted as a first order ODE since it is differential in only the $R_0$ variable) to obtain a global solution:

\begin{equation}
P_1 = \bar{P}_1(Z_0) R_0^{1-N^2}.
\end{equation}
Substituting this solution into Eqn.~\ref{P-eqn}, we obtain
 
\begin{equation}
\bar{P}_1^{\prime\prime}/\bar{P}_1 = N^2 (1 - N^2) R_0^{-2}.
\end{equation} 
The left hand of this equation does not depend on $R$ whereas, the right hand clearly does, unless it is zero, \ie $N = 0, 1$ (the $N = 0$ case is uninteresting).  Taking $\bar{P}_1^{\prime\prime} = 0$, we obtain

\begin{equation}
\bar{P}_1 = A_0 + A_1 Z_0.
\end{equation}
The first term corresponds to a translation in the $x$-$y$ plane, and the second term corresponds to a tilt, \ie the solution corresponds to a axi-symmetry preserving deformation.  The solutions for $\hat{R}_1$ and $\hat{Z}_1$, obtained from Eqns.~\ref{R1-eqn}-\ref{Z1-eqn}, are also consistent with this conclusion.  The numerical solution also confirms this analytic solution.  Therefore, the two solutions to the boundary value problem in fact satisfy the differential constraint globally, not just at the boundary, but correspond to trivial and uninteresting deformations.

\bibliographystyle{unsrtnat}
\bibliography{Quasisymmetry-near-axisymmetry-vacuum-jpp-arxiv}

\end{document}